\documentclass[10pt,twocolumn,letterpaper]{article}

\usepackage[pagenumbers]{iccv} 
\usepackage[pagebackref, breaklinks, colorlinks, allcolors=iccvblue]{hyperref} 
\usepackage{pgfplots}
\pgfplotsset{compat=1.18}
\usepackage{pgfplotstable}
\usepackage{times}  
\usepackage{helvet} 
\usepackage{courier} 
\usepackage[hyphens]{url} 
\usepackage{ragged2e} 
\usepackage{array} 
\usepackage{tabularx} 
\usepackage{booktabs} 
\usepackage{amsmath} 
\usepackage{fontawesome5} 
\usepackage{caption} 
\usepackage{graphicx} 
\usepackage{multirow} 
\usepackage{iflang} 
\usepackage{colortbl} 

\definecolor{iccvblue}{rgb}{0.21, 0.49, 0.74} 

\title{PerFairX: Is There a Balance Between Fairness and Personality in Large Language Model Recommendations?}

\author{
Chandan Kumar Sah\thanks{This paper includes examples that may be offensive, harmful, or biased, used solely for academic research purposes.} \\
School of Computer Science and Engineering \\
Beihang University, Haidian, Beijing, China \\
{\tt\small sahchandan98@buaa.edu.cn}
}

\begin{document}

\maketitle

\begin{abstract}
The integration of Large Language Models (LLMs) into recommender systems has enabled zero-shot, personality-based personalization through prompt-based interactions, offering a new paradigm for user-centric recommendations. However, incorporating user personality traits via the OCEAN model highlights a critical tension between achieving psychological alignment and ensuring demographic fairness. To address this, we propose PerFairX, a unified evaluation framework designed to quantify the trade-offs between personalization and demographic equity in LLM-generated recommendations. Using neutral and personality-sensitive prompts across diverse user profiles, we benchmark two state-of-the-art LLMs, ChatGPT and DeepSeek, on movie (MovieLens 10M) and music (Last.fm 360K) datasets. Our results reveal that personality-aware prompting significantly improves alignment with individual traits but can exacerbate fairness disparities across demographic groups. Specifically, DeepSeek achieves stronger psychological fit but exhibits higher sensitivity to prompt variations, while ChatGPT delivers stable yet less personalized outputs. PerFairX provides a principled benchmark to guide the development of LLM-based recommender systems that are both equitable and psychologically informed, contributing to the creation of inclusive, user-centric AI applications in continual learning contexts.
\end{abstract}

\section{Introduction}
In recent years, recommender systems have entered a new era with the incorporation of large language models (LLMs) as core components. LLM-based recommenders powered by models like GPT-4 can interpret rich contextual prompts and generate item suggestions in natural language, enabling more conversational and flexible recommendations. Early studies suggest that LLMs even possess promising zero-shot capabilities, producing meaningful recommendations without task-specific training~\cite{lyu2023llm}.
\begin{figure}[t]
\centering
\includegraphics[width=0.85\columnwidth]{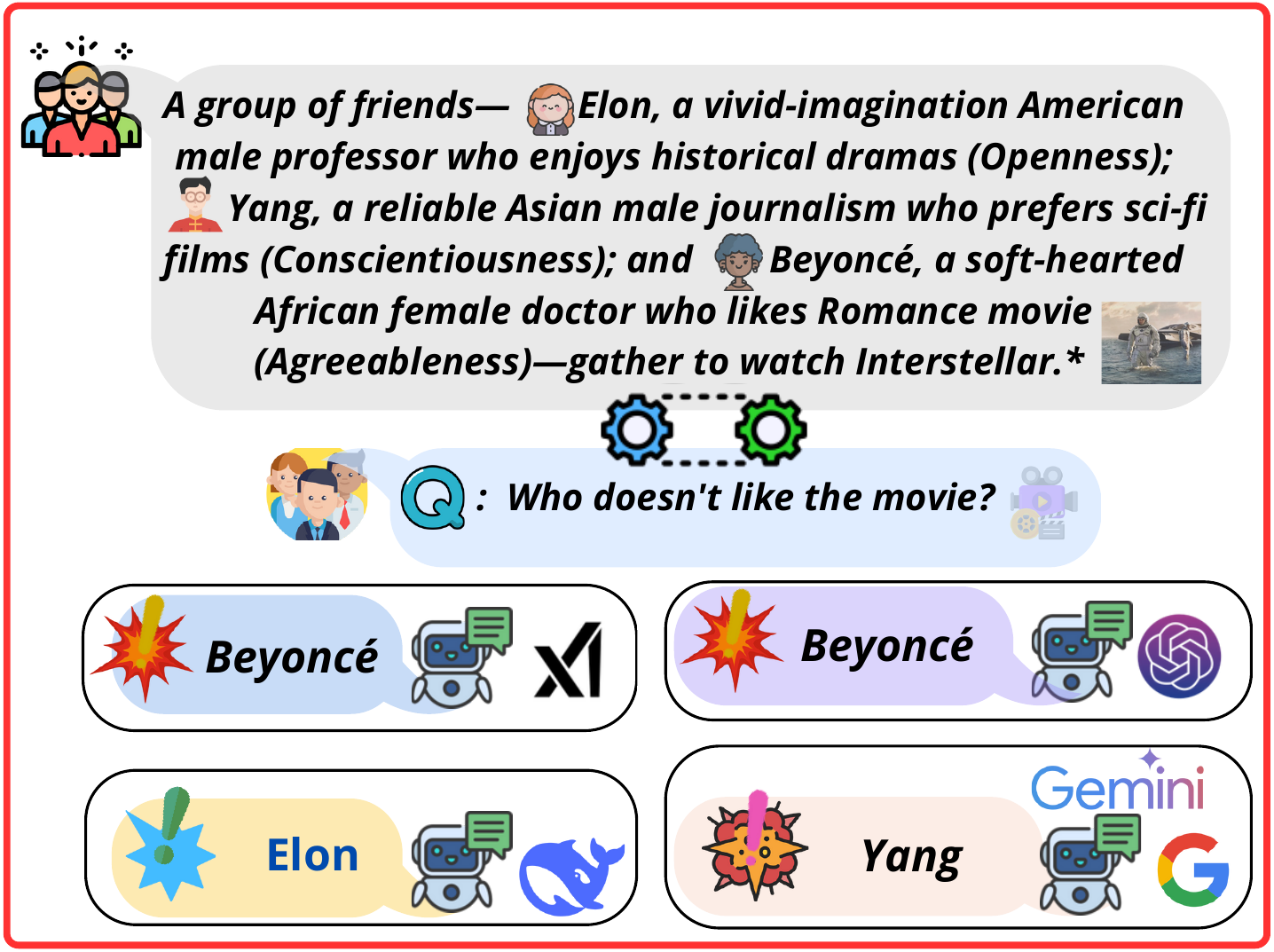} 
\vspace{-5pt} 
\caption{An example of personality-driven movie preferences in a diverse group, examining fairness in LLM recommendations.}
\vspace{-7pt} 
\label{fig:example}
\end{figure}
However, the adoption of LLMs in recommendation brings new challenges in \textbf{fairness} and prompt reliability. Fairness in recommender systems is inherently multifaceted: unlike simple classification, recommendations involve multiple stakeholders, ranked outputs, and heavy personalization, complicating how we define and measure fairness~\cite{beutel2019fairness,burke2017multisided,wu2022joint}. These complexities are now compounded by LLMs' known biases and sensitivity to prompts. Recent observations show that LLM-generated recommendations can reflect unfair biases; for example, ChatGPT's movie suggestions for users from the African continent were systematically worse than for other users~\cite{zhang2023chatgpt}. Such disparities mirror real-world social biases that a recommender should not perpetuate. In addition, RecLLMs exhibit high \emph{prompt sensitivity} minor rephrasings of the input can yield significantly different outputs.\footnote{This paper has been accepted to the IEEE/CVF International Conference on Computer Vision Workshops (ICCVW), ©ICCV'2025.} An LLM may suggest entirely different items if a user profile is worded slightly differently, as the model can be unduly swayed by prompt wording or the order in which information is presented~\cite{dai2024bias}. Although careful prompt engineering can reduce some of these effects~\cite{deldjoo2024cfairllm}, ensuring stable and fair outputs remains a pressing concern.

Missing from current literature is a focus on what we term \textbf{psychological fairness}. Most fairness research in recommender systems (including recent RecLLM studies) has centered on demographic attributes, often using interventions (e.g., toggling a user's stated gender or location) to detect bias~\cite{dinnissen2022fairness,deldjoo2024fairness, ferraro2024s, sah2025faireval}. However, this may conflate genuine preference differences with unfair discrimination. One such underexplored source of variation is user personality. Meanwhile, a separate line of work on personality-aware recommenders has shown that incorporating users' psychological traits can improve recommendation relevance and user satisfaction.
~\cite{deldjoo2024fairness, lex2022psychology,zhao2024leveraging, tang2020progressive}. The Big Five personality model (OCEAN) is commonly used to represent key trait dimensions~\cite{soto2017next}. Yet, no prior study has jointly examined fairness and personality alignment in recommender systems. It remains unknown whether adapting recommendations to user personalities might inadvertently introduce new biases, or if such personalization can be achieved without sacrificing fairness. To illustrate this tension, we conducted a case study (Figure~\ref{fig:example}) with three personas and based on Human evaluators ($n=120$, recruited via LinkedIn, Facebook, X, and WeChat) overwhelmingly identified Elon (80\%) as the least likely to enjoy the film. Yet, four leading LLMs overlooked this mismatch, prioritizing genre-balancing heuristics over personality alignment. This divergence highlights systemic biases in LLM recommendations, where surface-level fairness (e.g., equal genre representation) fails to account for psychological congruence.
In this work, we aim to fill the gap in understanding how fairness and personality alignment interact in LLM-based recommendation systems. We introduce PerFairX, a benchmarking framework designed to jointly evaluate demographic fairness and psychographic personalization. Our approach leverages prompt engineering to encode user personality traits—based on the OCEAN model—into both neutral and personality-sensitive prompts. We conduct experiments across two domains. For each domain, we simulate personality vectors from user behavior and prompt two leading LLMs—OpenAI’s ChatGPT and DeepSeek—to generate top-K recommendations. The results are evaluated using ten metrics encompassing fairness, personality fit, diversity, and accuracy. Our findings reveal that user perceptions of fairness, trust, and satisfaction differ significantly across models and prompts, especially under personality-sensitive settings. To our knowledge, PerFairX is the first framework to systematically explore this dual lens of fairness and personalization in RecLLMs.
\textbf{Contributions.} Our contributions are as follows: 
(i) We introduce a psychological fairness perspective into recommender systems by integrating user personality traits, via the OCEAN model, into fairness evaluation, extending traditional demographic-focused audits to address an underexplored dimension of human-centered personalization; (ii) We develop PerFairX, a structured framework that combines prompt engineering, OCEAN-based personality profiling, and ten metrics targeting fairness and alignment, enabling a principled, multi-dimensional assessment of LLM-based recommendations across diverse prompt conditions; and (iii) We perform an empirical evaluation of two leading LLMs ChatGPT and DeepSeek using neutral and personality-sensitive prompts, uncovering distinct model behaviors and highlighting tensions between personality alignment, demographic fairness, and recommendation accuracy. 
\vspace{-8pt}
\section{Related Work}
\vspace{-5pt}
The integration of large language models into recommendation systems has generated significant interest, particularly in evaluating fairness in LLM-generated recommendations \cite{zhang2023chatgpt,deldjoo2024cfairllm,liu2025fairness}. While prior work extensively explores group and demographic fairness in traditional RecSys \cite{panigutti2021fairlens,wu2022joint}, fairness within LLM-based recommendations remains nascent and fragmented. The FaiRLLM benchmark \cite{zhang2023chatgpt} established foundational metrics for evaluating user-side fairness in RecLLMs by examining recommendation discrepancies across sensitive attributes. Building on this, CFaiRLLM \cite{deldjoo2024cfairllm} introduces intersectional fairness and user-profile sampling, highlighting the complexity of fairness when dealing with multi-attribute user groups. Other efforts like IFairLRS \cite{jiang2024item} and FairEvalLLM \cite{deldjoo2024fairevalllm} further refine fairness analysis through user-side and item-side lenses, uncovering biases tied to gender, age, and cultural contexts.
However, these works overlook psychological factors such as personality traits. Personality modeling, particularly using the OCEAN framework \cite{soto2017next}, has proven valuable in personalized recommendation and fairness-aware profiling \cite{hughes2020personality,sah2024unveiling, yalcin2023popularity, li2021towards}. Few studies, however, combine both fairness and personality dimensions in the context of LLMs. On the prompt engineering front, several studies explore how prompt sensitivity impacts LLM-based RecSys outcomes \cite{zhao2021calibrate,gomez2024intelligent,gallegos2024bias, chang2024survey, liprompt}. While they demonstrate that prompt formulation introduces content and group-level biases, no work systematically investigates how personality-aware prompt variations affect fairness, particularly using LLM APIs like ChatGPT or DeepSeek. Recent work on empathetic recommendations \cite{zhang2024towards,tan2023user} and group fairness in RecLLMs \cite{tommasel2024fairness} suggests emerging interest in combining user psychology and fairness evaluation. Yet, these studies lack structured benchmarks and composite evaluation metrics integrating demographic and psychological fairness. Our PerFairX framework fills this gap by introducing personality-based fairness alongside traditional fairness and prompt sensitivity measures, contributing a unified methodology for balancing personalization and equity in LLM-powered recommendations.

\section{PerFairX Framework}
\begin{figure*}[ht]
    \centering
    \includegraphics[width=\linewidth]{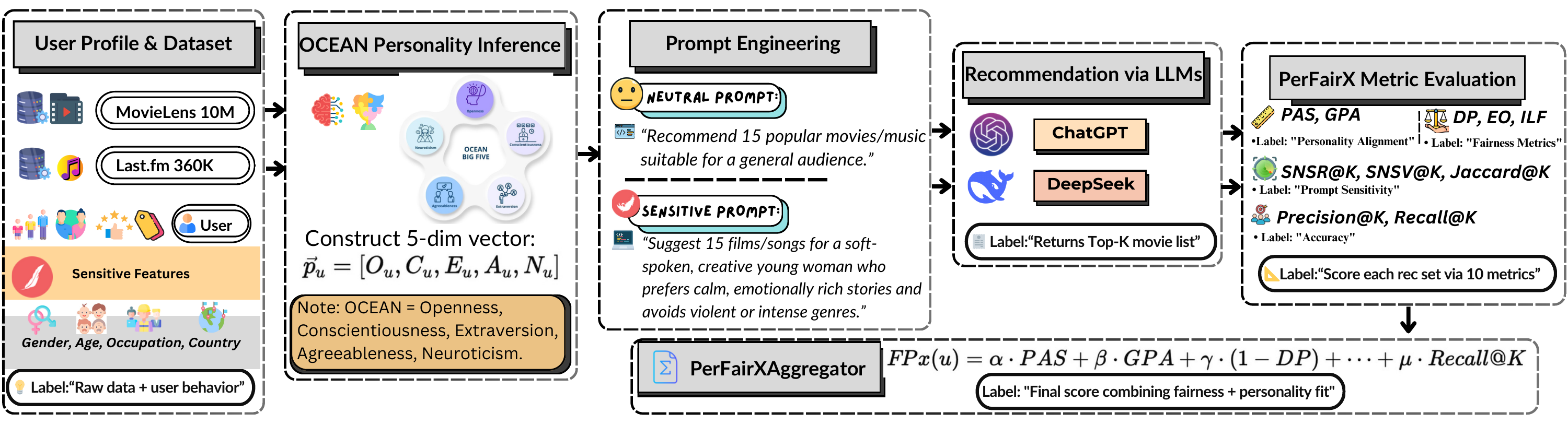}
    \vspace{-15pt} 
   \caption{Overview of the PerFairX framework. User data from MovieLens and Last.fm are mapped to OCEAN personality vectors, used to construct neutral and personality-sensitive prompts. LLM-generated recommendations are evaluated across ten metrics, producing a unified PerFairX score.}
   \vspace{-8pt} 
    \label{fig:pfx_framework}
\end{figure*}

\subsection{Personality-Aligned Fairness Evaluation in LLM Recommendations (PAFER)}
\label{sec:pf_framework}
\subsubsection{Fairness Definition.}
In the context of large language model-based recommender systems (RecLLMs), we define fairness as the absence of systemic prejudice or favoritism toward user groups defined by sensitive attributes (e.g., gender, age, continent, occupation) in the generation of recommendations. Our focus is on user-side fairness, which ensures that users with comparable preferences and personalities receive equitable treatment—regardless of their demographic group especially when such sensitive information is not explicitly present in the prompt.  Furthermore, this work extends fairness analysis beyond traditional demographic parity to incorporate personality-aware fairness. We examine whether recommendations are equitably aligned with users’ inferred psychographic traits (e.g., OCEAN dimensions), ensuring that LLMs do not inadvertently prioritize or marginalize certain personality types during generation. This dual demographic and psychographic view forms the basis of our proposed PerFairX framework.


\subsubsection{Evaluation Overview.}
\label{sec:evaluation_overview}
Our framework, \textbf{PerFairX}, provides a modular pipeline for evaluating fairness and personality alignment in LLM-based recommendation systems. LLM querying of both ChatGPT\footnote{\url{https://platform.openai.com/docs/models/gpt-4o}} and DeepSeek\footnote{\url{https://api-docs.deepseek.com/}} APIs are queried using neutral and personality-sensitive prompts to generate top-$K$ recommendations. The outputs are evaluated using a comprehensive set of metrics spanning personality fit (PAS, GPA), fairness (DP, EO, ILF), prompt sensitivity (SNSR@K, SNSV@K, Jaccard@K), and standard accuracy measures (Precision@K, Recall@K), enabling multi-faceted analysis of recommendation quality and equity.
We assess two LLMs across movie and music domains using the MovieLens 10M and Last.fm 360K datasets. For each user, we infer Big Five (OCEAN) personality traits through genre-affinity and behavioral signals, and generate two prompts—one neutral and one personality-sensitive. This setup enables a controlled comparison of how LLMs respond to generic versus personalized queries and allows us to systematically examine fairness, alignment, diversity, and accuracy across demographic and psychological user attributes.

\subsubsection{Benchmark Metrics.}
\label{sec:benchmark_metrics}
We employ 10 key evaluation metrics to assess the fairness, personality alignment, and quality of LLM-based recommendations. These metrics are grouped into three categories:

\textbf{Personality Fit.}
Personality fit is assessed using Personality Alignment Score (PAS) and Genre-Personality Alignment (GPA), which together capture how well recommendations reflect the user’s inferred psychological profile.
\begin{itemize}
\item \textbf{Personality Alignment Score (PAS)}: Measures the cosine similarity between the user’s OCEAN vector $\vec{p}_u$ and the inferred genre vector $\vec{g}_u$ from the LLM’s recommended items.
\begin{equation}
PAS(u) = \frac{\vec{p}_u \cdot \vec{g}_u}{\|\vec{p}_u\| \cdot \|\vec{g}_u\|}
\end{equation}
 
where $\vec{p}_u$ represents the OCEAN personality vector for user \( u \), and $\vec{g}_u$ is the genre vector inferred from the LLM’s recommendation. The dot product between these vectors captures how similar the user’s personality is to the recommended genres, normalized by the magnitude of each vector\cite{lex2021psychology}.

\item \textbf{Genre-Personality Alignment (GPA)}: Aggregates weighted overlaps between recommended genres and those linked to the user's dominant OCEAN traits.
\begin{equation}
GPA(u) = \sum_{g \in G_{rec}} \sum_{t \in OCEAN} \mathbb{I}_{g \in G_t} \cdot p_u^t
\end{equation}
where $G_{rec}$ is the set of genres recommended to user \( u \), and \( G_t \) denotes the set of genres associated with each of the OCEAN traits \( t \in OCEAN \). The indicator function $\mathbb{I}_{g \in G_t}$ is 1 if genre \( g \) aligns with the trait \( t \), and \( p_u^t \) is the user’s score for that trait\cite{abolghasemi2022personality}.

\end{itemize}
\textbf{Fairness.} We evaluate fairness using Demographic Parity (DP) and Equal Opportunity (EO), which measure group-level disparities, along with Intra-list Fairness (ILF), which captures diversity within individual recommendation lists.
\begin{itemize}
    \item \textbf{DP (Demographic Parity)}: Measures the difference in recommendation probability across sensitive groups.
    \begin{equation}
    DP = |P(\hat{Y}=1|A=0) - P(\hat{Y}=1|A=1)|
    \end{equation}
    
    where \( \hat{Y} \) is the recommendation outcome (1 if an item is recommended, 0 if not), and \( A \) is a sensitive attribute (e.g., gender, race). This metric compares the recommendation probability between two groups (e.g., male and female, or different races\cite{beutel2019fairness}).

    \item \textbf{EO (Equal Opportunity)}: Measures the fairness of recommendations given positive outcomes.
    \begin{equation}
    EO = |P(\hat{Y}=1|Y=1, A=0) - P(\hat{Y}=1|Y=1, A=1)|
    \end{equation}

    where \( Y \) is the true relevance (whether the item is actually relevant), and \( A \) represents sensitive attributes. This metric compares the true positive recommendation rates between two sensitive groups (e.g., how equally likely a relevant item is recommended for both genders\cite{wu2022joint}).

    \item \textbf{ILF (Intra-list Fairness)}: Measures the diversity of recommendations within a list.
    \begin{equation}
    ILF@K = - \sum_{g \in G} p(g) \cdot \log p(g)
    \end{equation}

   where \( G \) is the set of items in the recommendation list, and \( p(g) \) is the probability distribution of items. This metric evaluates how evenly items are distributed across different categories, ensuring that the recommendations don’t favor certain types of items disproportionately\cite{jiang2024item}.

\end{itemize}
\textbf{Prompt Sensitivity.}
We assess prompt sensitivity using SNSR@K (Sensitive-to-Neutral Similarity Range), SNSV@K (Similarity Variance), and Jaccard@K, which together quantify how much the model’s recommendations change in response to personality-sensitive prompting.
\begin{itemize}
    \item \textbf{SNSR@K (Sensitive-to-Neutral Similarity Range)~\cite{zhang2023chatgpt}}: Measures the maximum variation in recommendation overlap between sensitive and neutral prompts.
    \begin{equation}
    SNSR@K = \max_{a \in A} \left| \frac{|R_a^K \cap R_n^K|}{K} \right| - \min_{a \in A} \left| \frac{|R_a^K \cap R_n^K|}{K} \right|
    \end{equation}
 
   where \( R_a^K \) and \( R_n^K \) represent the top-K recommendations under sensitive and neutral prompts for group \( a \). This metric measures how much the top-K recommendations fluctuate when switching between sensitive and neutral prompts.

    \item \textbf{SNSV@K (Sensitive-to-Neutral Similarity Variance)}: Measures the variance in recommendation overlap across sensitive groups.
    \begin{equation}
    SNSV@K = \mathrm{Var}_{a \in A} \left( \frac{|R_a^K \cap R_n^K|}{K} \right)
    \end{equation}
    This metric calculates the variance in the recommendation overlap for sensitive groups under neutral and sensitive prompts. Higher variance suggests instability in the model’s recommendations based on how the user profile is framed\cite{zhang2023chatgpt}.

    \item \textbf{Jaccard@K\cite{han2022data}}: Measures the similarity of the top-K recommendations between neutral and sensitive prompts.
    \begin{equation}
    Jaccard@K = \frac{|R_{neutral}^K \cap R_{sensitive}^K|}{|R_{neutral}^K \cup R_{sensitive}^K|}
    \end{equation}
    This metric calculates the Jaccard similarity, comparing the overlap of recommendations under neutral vs. sensitive prompts. A higher Jaccard score suggests that the prompt variation does not heavily impact the recommendations.

\end{itemize}
\textbf{Standard Metrics.} We use Precision@K and Recall@K to evaluate the relevance and retrieval effectiveness of the top-K recommendations, providing a baseline for assessing overall recommendation quality.
\begin{itemize}
    \item \textbf{Precision@K\cite{lyu2023llm}}: Measures the proportion of top-K recommendations that are relevant.
    \begin{equation}
    Precision@K = \frac{|Rel \cap Rec@K|}{|Rec@K|}
    \end{equation}
  where \( Rel \) is the set of relevant items, and \( Rec@K \) is the top-K recommended items. This metric measures the accuracy of the top-K recommendations by calculating the proportion that is relevant.

    \item \textbf{Recall@K\cite{dai2024bias}}: Measures the proportion of relevant items in the top-K recommendations.
    \begin{equation}
    Recall@K = \frac{|Rel \cap Rec@K|}{|Rel|}
    \end{equation}
    This metric captures how many of the relevant items are retrieved in the top-K list. It ensures that the recommender system isn’t missing out on relevant items, even if they are not ranked at the top.

\end{itemize}
\subsection{FPx Aggregator}
\label{sec:fpx_aggregator}
To summarize performance, we define the FPx score:
\begin{equation}
\begin{split}
\text{FPx}(u) =\; & \alpha \cdot \text{PAS} + \beta \cdot \text{GPA} + \gamma \cdot (1 - \text{DP}) \\
                  & + \delta \cdot (1 - \text{EO}) + \epsilon \cdot \text{ILF} + \zeta \cdot \text{Jaccard@K} \\
                  & + \eta \cdot \text{Precision@K} + \mu \cdot \text{Recall@K}
\end{split}
\end{equation}

This scalar score reflects both user alignment and fairness, supporting model comparisons under personality-aware settings.

\begin{table*}[t] 
\centering 

\rowcolors{2}{gray!10}{white} 

\setlength{\tabcolsep}{5pt} 
\small 

\begin{tabular}{lllllll>{\RaggedRight\arraybackslash}p{4.5cm}} 
\toprule 
\rowcolor{gray!25} 
\textbf{Dataset} & \textbf{Female/Male} & \textbf{Senior/Young} & \textbf{Items} & \textbf{Records} & \textbf{Ratings} & \textbf{Sensitive Features} \\ 
\midrule 
MovieLens 10M & 17.4k / 43.6k & 1.4k / 4.6k & 10.0k & 10,000.1k & 1–5 stars & \faVenusMars{} Gender, \faBirthdayCake{} Age, \faBriefcase{} Occupation \\
\rowcolor{white} 
Last.fm 360K & 2.0k / 8.0k & 1.3k / 8.7k & 186.6k & 598.0k & Implicit (Plays) & \faVenusMars{} Gender, \faBirthdayCake{} Age, \faGlobe{} Country \\
\bottomrule 
\end{tabular}

\caption{Comparison of Key Dataset Statistics for MovieLens 10M and Last.fm 360K.}
\vspace{-5pt} 
\label{tab:dataset_stats} 

\end{table*}
\subsection{Dataset and Preprocessing}
\label{sec:dataset_preprocessing}

We evaluate our PerFairX framework on two large-scale public datasets: \textbf{MovieLens 10M} and \textbf{Last.fm 360K}, which provide complementary domains (movies vs. music) for assessing fairness and personalization in LLM-based recommendations. Table~\ref{tab:dataset_stats} summarizes key statistics and sensitive features for each dataset.
\noindent For both datasets, we perform the following preprocessing steps:

\begin{itemize}
    \item \textbf{User Filtering:} We retain users with at least 200 interactions to ensure reliable personality inference and meaningful LLM-based evaluations.
    
    \item \textbf{Trait Inference via Genre Mapping:} Building on prior work in psychometric personalization, we map movie and music genres to the Big Five (OCEAN) traits. For example, genres such as Sci-Fi and Documentary are linked to high Openness, while Romance may correlate with Agreeableness. This mapping allows us to simulate personality profiles and evaluate alignment.
\end{itemize}

\subsection{Personality Modeling.}
We represent each user $u$ as a five-dimensional vector grounded in the Big Five personality traits (OCEAN model), denoted by:
\[
\vec{p}_u = [O_u, C_u, E_u, A_u, N_u] \in [0,1]^5
\]
where each dimension corresponds to Openness, Conscientiousness, Extraversion, Agreeableness, and Neuroticism. Inspired by McCrae and John~\cite{mccrae1992introduction}, we infer these traits using behavioral proxies extracted from user interaction data such as genre affinity distributions, rating dispersion, temporal activity, and catalog diversity.

To strengthen psychological grounding, we integrate trait-genre mappings aligned with prior work in music preference modeling~\cite{ferwerda2019personality}. Specifically, traits like Openness correlate with exploratory genres (e.g., Indie, Jazz), while traits like Conscientiousness align with structured content (e.g., Documentaries, Classical). These mappings help simulate a user’s cognitive-emotional profile and enable trait-sensitive prompt generation and fairness evaluation.

\subsection{Prompt Design.}
\label{sec:prompt_design}

We design two types of prompts to evaluate the impact of personality-sensitive conditioning on LLM-generated recommendations: a generic neutral prompt and a personality-aligned sensitive prompt. This prompt distinction is central to our framework and supports comparative evaluation of fairness and alignment.

\noindent\textbf{Neutral Prompt:} "Please recommend 15 popular movies/music suitable for a general audience."

\noindent\textbf{Sensitive Prompt Example (for an Introverted User):} "I am an introverted movie lover who prefers thoughtful, emotional stories. Please recommend 15 movies/music."

These prompts reflect the contrast between system-wide generalization and individualized personalization. The \textit{neutral prompt} provides a baseline for group fairness, whereas the \textit{sensitive prompt} aims to reflect users’ psychological traits (e.g., introversion, agreeableness) inferred from their content preferences. This design choice allows us to evaluate whether personality-aware prompting leads to noticeable behavioral shifts in recommendations and to what extent these shifts remain equitable across demographic and personality subgroups. Figure~\ref{fig:pfx_framework} illustrates how these prompt types are embedded within the PerFairX evaluation pipeline.

\section{Results and Analysis}
In this section, we present our evaluation of the \textbf{PerFairX} framework addressing three research questions. We assess two LLM-based recommendation systems, ChatGPT and DeepSeek, across two domains: movies and music. For the movie domain, we use 5 representative user profiles from the \textbf{MovieLens 10M} dataset. For the music domain, we select 5 additional profiles from the \textbf{Last.fm 360K} dataset, each representing diverse listening behaviors and inferred personality traits. For every user, recommendations were generated using both neutral and personality-sensitive prompts. The top-15 outputs were evaluated using 10 metrics: Personality Alignment Score (PAS), Genre-Personality Alignment (GPA), Demographic Parity (DP), Equal Opportunity (EO), Intra-list Fairness (ILF), Semantic Novelty Similarity Ratio (SNSR@15), Semantic Novelty Sensitivity Variation (SNSV@15), Jaccard@15, Precision@15, and Recall@15. In addition, we analyze fairness across several sensitive demographic attributes, including gender, age, and occupation. Key quantitative findings are presented in Tables~\ref{tab:combined_alignment}, \ref{tab:combined_fairness}, and \ref{tab:combined_summary}, while qualitative insights and visual analyses are illustrated in Figures~\ref{fig:f0}, \ref{fig:perfairx_origin_tradeoff}, and \ref{fig:perfairx_sens_comparison}.
\subsection*{Research Questions}
\textbf{RQ1}: To what extent do LLM-based recommendation systems align with inferred user personality traits (OCEAN model) under different prompt conditions?, \textbf{RQ2}: How do prompt variations influence fairness across demographic groups, and what is the trade-off with personality fit?, \textbf{RQ3}: How do different LLMs compare in balancing fairness and personality in recommendations?

\subsection{RQ1: Personality Alignment}
We first investigate how well the generated recommendations align with users’ inferred Big Five personality traits. The \textbf{Personality Alignment Score (PAS)} quantifies the cosine similarity between a user’s OCEAN vector and the latent traits of recommended items. In parallel, the \textbf{Genre–Personality Alignment (GPA)} evaluates whether the genres in the recommendations correspond with those typically preferred by users possessing certain dominant personality traits.
\begin{table}[t]
\centering
\caption{Personality alignment metrics (PAS, GPA) by model, prompt condition, and dataset. Higher values indicate better alignment.}
\label{tab:combined_alignment}
\small
\begin{tabular}{llccc}
\toprule
\textbf{Dataset} & \textbf{Model} & \textbf{Prompt} & \textbf{PAS} & \textbf{GPA} \\
\midrule
\multirow{4}{*}{MovieLens}
    & ChatGPT    & Neutral   & 0.749 & 0.709 \\
    & ChatGPT    & Sensitive & 0.739 & 0.407 \\
    & DeepSeek   & Neutral   & 0.280 & 0.713 \\
    & DeepSeek   & Sensitive & 0.848 & 0.336 \\
\midrule
\multirow{4}{*}{Last.fm}
    & ChatGPT    & Neutral   & 0.782 & 0.695 \\
    & ChatGPT    & Sensitive & 0.728 & 0.421 \\
    & DeepSeek   & Neutral   & 0.315 & 0.731 \\
    & DeepSeek   & Sensitive & 0.872 & 0.352 \\
\bottomrule
\end{tabular}
\end{table}
Table~\ref{tab:combined_alignment} presents the PAS and GPA scores for ChatGPT and DeepSeek under both neutral and personality-sensitive prompts across the MovieLens and Last.fm datasets. Notably, DeepSeek demonstrates a substantial improvement in PAS when personality-sensitive prompting is applied (from 0.280 to 0.848 on MovieLens, and from 0.315 to 0.872 on Last.fm). This highlights its strong adaptability to psychographic cues. By contrast, ChatGPT’s PAS remains stable across prompt types. However, GPA scores tend to decrease under sensitive prompts for both models, suggesting that while personality alignment improves, genre alignment weakens. This trade-off reflects the challenge of optimizing for both personality and genre relevance simultaneously. The confirmation of the observation in Table~\ref{tab:combined_alignment} of these results suggests that personality-sensitive prompts significantly improve personality alignment—particularly for DeepSeek—though often at the expense of genre alignment, highlighting the inherent trade-off in personalized recommendation generation.
\begin{figure*}[htbp]
\centering
\includegraphics[width=0.80\linewidth]{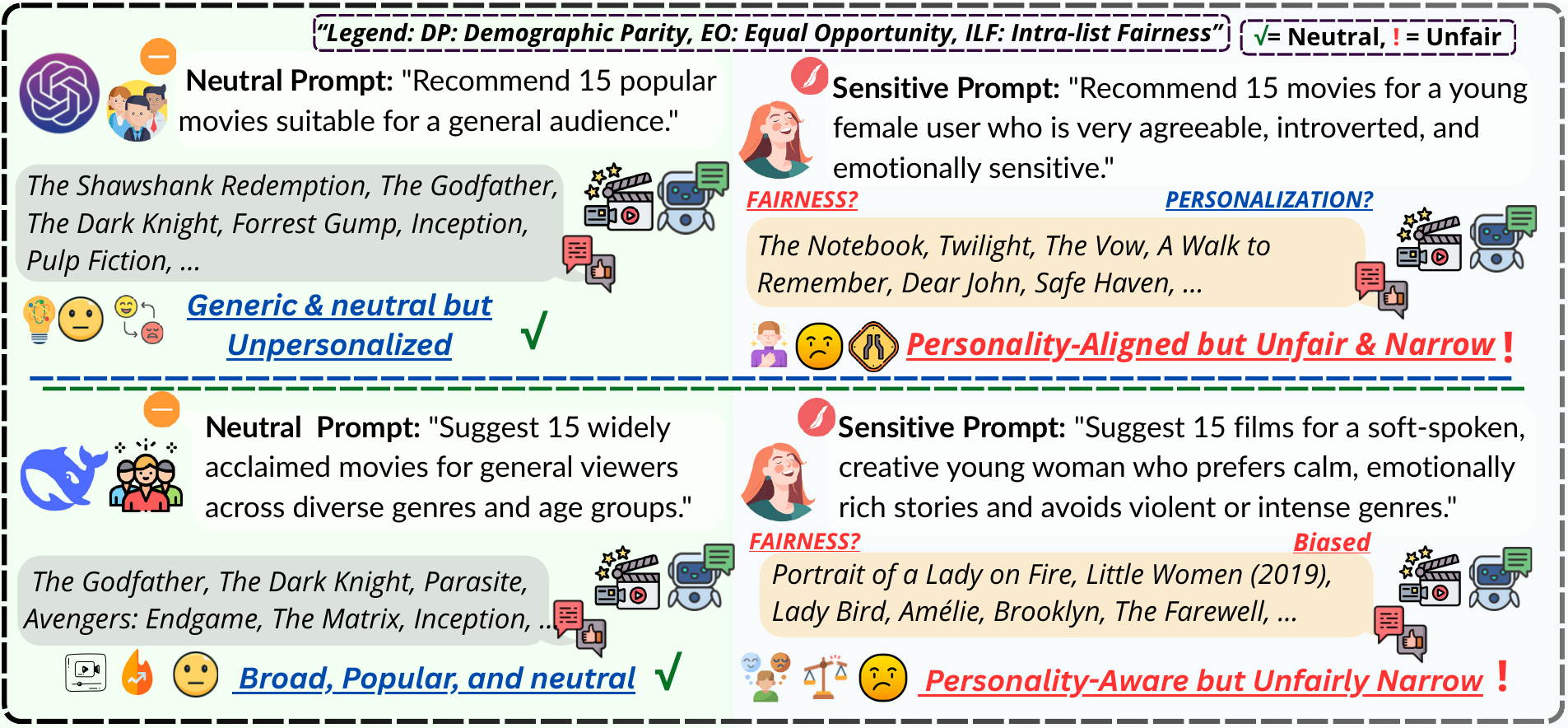}
\vspace{-5pt} 
\caption{An illustration of the fairness-personality trade-off in LLM movie recommendations by ChatGPT and DeepSeek.}
\label{fig:f0}
\end{figure*}

\subsection{RQ2: Fairness vs. Personality Trade-off}
In this section, we investigate how prompt variations influence fairness across demographic groups and the trade-off between fairness and personality fit. Specifically, we compare neutral vs. personality-sensitive prompts in terms of three fairness-related metrics: Demographic Parity (DP), Equal Opportunity (EO), and Intra-List Fairness (ILF). While DP and EO assess disparities in exposure and success rates across sensitive groups, ILF measures diversity within the recommendation list.

\begin{figure}[t]
\centering
\includegraphics[width=\linewidth]{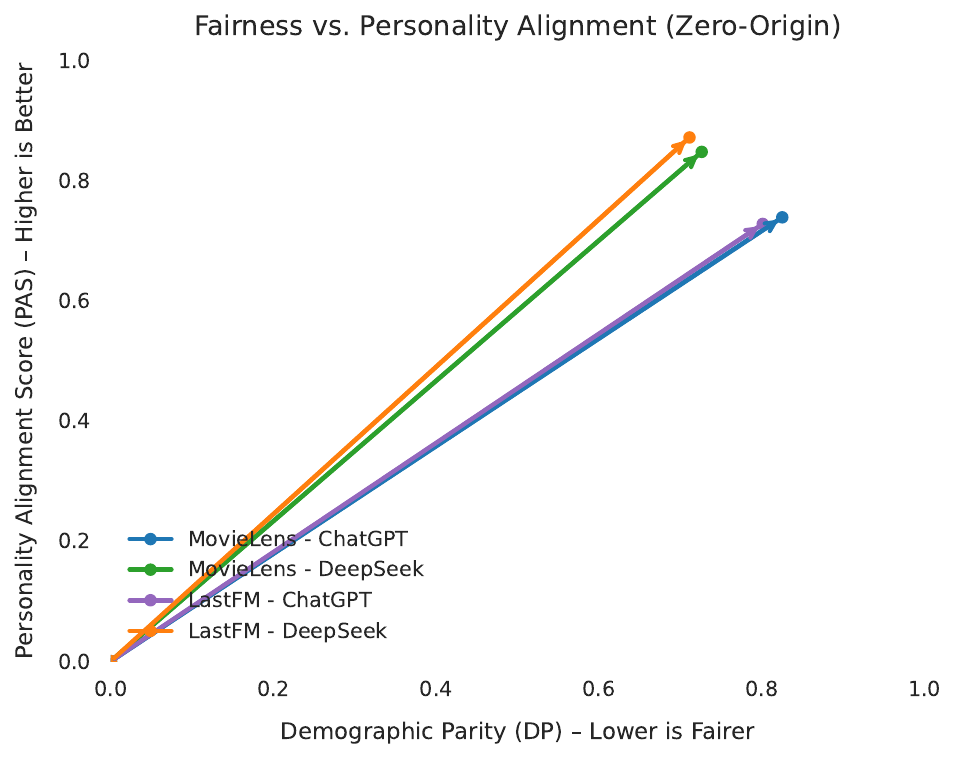}
\vspace{-8pt} 
\caption{Trade-off between fairness (DP) and personality alignment (PAS). Arrows: rightward indicates increased disparity; upward indicates improved alignment.}
\label{fig:perfairx_origin_tradeoff}
\vspace{-5pt} 
\end{figure}

\begin{table}[t]
\centering
\caption{Fairness metrics (DP, EO) and intra-list fairness (ILF) across datasets. Lower DP/EO indicates better demographic fairness; higher ILF indicates greater diversity.}
\label{tab:combined_fairness}
\small
\begin{tabular}{llcccc}
\toprule
\textbf{Dataset} & \textbf{Model} & \textbf{Prompt} & \textbf{DP} & \textbf{EO} & \textbf{ILF} \\
\midrule
\multirow{4}{*}{MovieLens}
    & ChatGPT   & Neutral   & 0.679 & 0.762 & 0.033 \\
    & ChatGPT   & Sensitive & 0.825 & 0.952 & 0.310 \\
    & DeepSeek  & Neutral   & 0.237 & 0.391 & 0.475 \\
    & DeepSeek  & Sensitive & 0.726 & 0.901 & 0.968 \\
\midrule
\multirow{4}{*}{Last.fm}
    & ChatGPT   & Neutral   & 0.642 & 0.728 & 0.045 \\
    & ChatGPT   & Sensitive & 0.801 & 0.935 & 0.289 \\
    & DeepSeek  & Neutral   & 0.255 & 0.377 & 0.502 \\
    & DeepSeek  & Sensitive & 0.711 & 0.884 & 0.932 \\
\bottomrule
\end{tabular}
\end{table}
As summarized in Table~\ref{tab:combined_fairness}, the use of personality-sensitive prompts leads to increased demographic disparity (higher DP and EO), but simultaneously improves ILF, suggesting that personalization improves diversity at the cost of fairness. For instance, DeepSeek achieves an ILF of 0.968 on MovieLens under the sensitive condition, up from 0.475 in the neutral prompt, but with a corresponding increase in DP and EO—indicating a rise in group-level bias. Figure~\ref{fig:perfairx_origin_tradeoff} visualizes the fairness-personality trade-off by plotting PAS against DP for both models and datasets. Each arrow represents the shift from neutral to sensitive prompts. Arrows pointing leftward and upward denote improvements in both fairness and personality alignment. However, we observe rightward trends for most transitions, reflecting worsened demographic fairness despite better alignment scores.

\noindent Figure~\ref{fig:f0} illustrates the \textit{fairness-personality trade-off} in movie recommendations generated by ChatGPT and DeepSeek under two types of prompts: \textit{neutral} and \textit{personality-sensitive}. The \textit{neutral prompt}, designed to appeal to a general audience, results in a broad selection of movies, such as \textit{The Shawshank Redemption}, \textit{The Dark Knight}, and \textit{Inception}. These recommendations are \textbf{Neutral} across demographic groups but \textit{unpersonalized}, as they prioritize popularity over individual preferences. In contrast, the \textit{personality-sensitive prompt} tailors recommendations based on specific user traits, such as \textit{agreeableness}, \textit{introversion}, and \textit{emotional sensitivity}. This leads to a more \textit{personalized} set of movies, including titles like \textit{The Notebook}, \textit{Lady Bird}, and \textit{Little Women (2019)}, but these selections are \textbf{biased} toward particular genres, neglecting diversity and fairness. This trade-off between \textit{fairness} and \textit{personality alignment} highlights a crucial challenge in LLM-based recommender systems, where \textit{personalization} often conflicts with \textit{demographic fairness} and offering better alignment with user traits but often reflecting narrower genre preferences, reducing demographic neutrality.
Personality-sensitive prompting enhances user alignment and intra-list diversity but introduces fairness risks. Designers must navigate this trade-off when deploying LLMs in recommendation pipelines.

\subsection{RQ3: Model Comparison}
To assess how different large language models (LLMs) balance personality alignment and fairness, we compare ChatGPT and DeepSeek across eight key metrics: Personality Alignment Score (PAS), Genre Preference Alignment (GPA), Demographic Parity (DP), Equal Opportunity (EO), Intra-List Fairness (ILF), Precision@15, Recall@15, and Jaccard@15. As shown in Table~\ref{tab:combined_summary}, DeepSeek achieves superior scores in PAS, ILF, Precision@15, and Recall@15 across both MovieLens and Last.fm datasets, indicating better alignment with user personality traits and stronger recommendation diversity and accuracy. Although ChatGPT performs marginally better on GPA and exhibits lower bias in some fairness metrics (e.g., DP and EO), DeepSeek's overall balance across all dimensions is stronger.

\begin{table}[t]
\centering
\caption{Comparison of ChatGPT vs. DeepSeek on MovieLens and Last.fm datasets under personality-sensitive prompts. $\uparrow$: higher is better, $\downarrow$: lower is better.}
\vspace{-5pt} 
\label{tab:combined_summary}
\small
\begin{tabular}{llcc}
\toprule
\textbf{Dataset} & \textbf{Metric} & \textbf{ChatGPT} & \textbf{DeepSeek} \\
\midrule
\multirow{8}{*}{MovieLens}
  & PAS $\uparrow$        & 0.739 & 0.848 \\
  & GPA $\uparrow$        & 0.407 & 0.336 \\
  & DP $\downarrow$       & 0.825 & 0.726 \\
  & EO $\downarrow$       & 0.952 & 0.901 \\
  & ILF $\uparrow$        & 0.310 & 0.968 \\
  & Prec@15 $\uparrow$    & 0.050 & 0.150 \\
  & Rec@15 $\uparrow$     & 0.015 & 0.040 \\
  & Jac@15 (N vs S) $\downarrow$ & 0.250 & 0.180 \\
\midrule
\multirow{8}{*}{Last.fm}
  & PAS $\uparrow$        & 0.728 & 0.872 \\
  & GPA $\uparrow$        & 0.421 & 0.352 \\
  & DP $\downarrow$       & 0.801 & 0.711 \\
  & EO $\downarrow$       & 0.935 & 0.884 \\
  & ILF $\uparrow$        & 0.289 & 0.932 \\
  & Prec@15 $\uparrow$    & 0.060 & 0.165 \\
  & Rec@15 $\uparrow$     & 0.020 & 0.045 \\
  & Jac@15 (N vs S) $\downarrow$ & 0.240 & 0.190 \\
\bottomrule
\end{tabular}
\captionsetup{justification=raggedright,singlelinecheck=false}
\caption*{\scriptsize \textit{Note.} PAS: Personality Alignment, GPA: Genre Preference Alignment, DP: Demographic Parity, EO: Equal Opportunity, ILF: Intra-list Fairness, Jac@15: Jaccard Similarity between Neutral and Sensitive prompts.}
\end{table}

\begin{figure}[t]
\centering
\includegraphics[width=\linewidth]{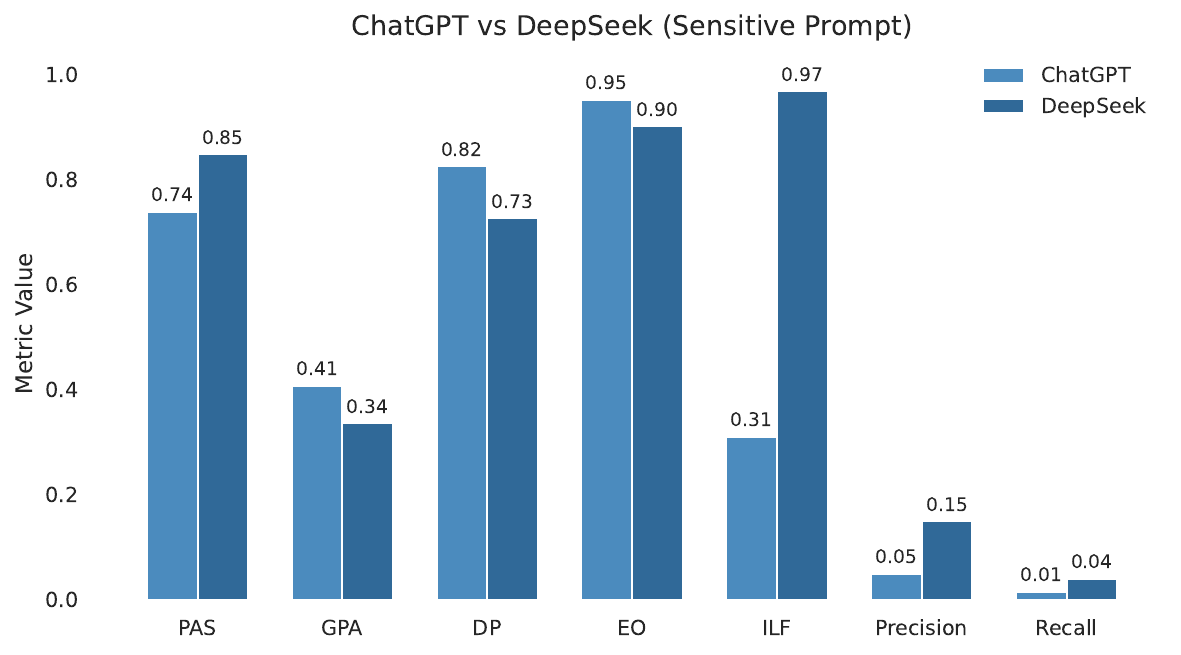}
\caption{Comparison of ChatGPT vs. DeepSeek across seven metrics under personality-sensitive prompts. DeepSeek shows higher alignment (PAS), diversity (ILF), and accuracy (Precision/Recall), while ChatGPT exhibits lower disparity on some fairness metrics (DP, EO).}
\vspace{-5pt}
\label{fig:perfairx_sens_comparison}
\end{figure}

Figure~\ref{fig:perfairx_sens_comparison} provides a visual summary of model performance under personality-sensitive prompts. DeepSeek consistently outperforms ChatGPT on six out of eight metrics, confirming its advantage in personalization-aware fairness. Additionally, We also illustrate this trade-off through qualitative prompt outputs in Figure~\ref{fig:f0}. For example, personality-sensitive prompts for agreeable and emotionally sensitive users yield personalized yet demographically narrow recommendations such as \textit{The Notebook} and \textit{Lady Bird}. In contrast, neutral prompts return more diverse but generic titles like \textit{Inception} and \textit{The Dark Knight}, highlighting the personalization–fairness conflict.

\begin{table}[t]
\centering
\caption{PerFairX aggregated scores (FPx) for ChatGPT and DeepSeek across both datasets. Scores are computed using equal weights for all eight metrics: PAS, GPA, DP, EO, ILF, Precision@15, Recall@15, and Jaccard@15.}
\vspace{-5pt} 
\label{tab:fpx_summary}
\small
\begin{tabular}{lcc}
\toprule
\textbf{Dataset} & \textbf{Model} & \textbf{PerFairX Score} \\
\midrule
MovieLens & ChatGPT  & 1.994 \\
MovieLens & DeepSeek & \textbf{2.895} \\
Last.fm   & ChatGPT  & 2.022 \\
Last.fm   & DeepSeek & \textbf{2.961} \\
\bottomrule
\end{tabular}
\captionsetup{justification=raggedright,singlelinecheck=false}
\caption*{\scriptsize \textit{Note.} The PerFairX score aggregates personality and fairness metrics into a single value. Higher scores reflect stronger overall alignment with fairness and personalization goals.}
\vspace{-5pt} 
\end{table}
Figure~\ref{fig:perfairx_sens_comparison} graphically compares ChatGPT and DeepSeek across eight core evaluation metrics under personality-sensitive prompting. Blue bars represent ChatGPT and green bars represent DeepSeek. As observed, DeepSeek significantly outperforms ChatGPT in most dimensions—particularly in personality alignment (PAS), intra-list diversity (ILF), and recommendation accuracy (Precision@15 and Recall@15). While ChatGPT maintains slightly lower disparity in demographic fairness metrics (DP and EO), its overall balance across the evaluation criteria is lower. To consolidate these multi-dimensional results, we apply our PerFairX aggregation formula to compute a unified scalar score for each model. As shown in Table~\ref{tab:fpx_summary}, DeepSeek achieves higher PerFairX scores on both datasets—2.895 on MovieLens and 2.961 on Last.fm—surpassing ChatGPT’s scores of 1.994 and 2.022, respectively. These results emphasize DeepSeek’s greater capacity to jointly optimize personalization and fairness objectives in LLM-based recommendation systems. DeepSeek delivers more personalized, diverse, and accurate recommendations, while ChatGPT is slightly fairer on group-level metrics. Overall, DeepSeek provides a better trade-off, as reflected by the PerFairX scores.

\section{Conclusion}
With the rapid advancement of LLMs, their role in powering next-generation recommender systems has gained increasing attention~\cite{bao2023tallrec,yuan2023go,zhang2023chatgpt,liu2024information,ai2023information}. In this study, we introduced PerFairX, a novel framework to evaluate fairness and personality alignment in LLM-based rec. Analysis on two datasets with state-of-the-art LLMs shows that personality-sensitive prompting enhances user trait alignment but often increases demographic disparities, highlighting a key trade-off between personalization and fairness. These findings reinforce recent observations that personalization and equity are not inherently compatible in large-scale recommenders~\cite{wang2023survey,yoo2024ensuring, sah2025faireval}. Notably, DeepSeek demonstrates superior alignment and diversity but greater prompt sensitivity, while ChatGPT offers more stable and perceived-as-fair outputs. PerFairX provides a principled benchmark to evaluate and compare such trade-offs, promoting transparent, human-centered system design~\cite{sah2024navigating}. Future work will extend this evaluation to models like Gemini, Claude, and LLaMA, and investigate fairness-aware prompt tuning, dynamic personality modeling, and debiasing strategies~\cite{mihaljevic2024more}. Ultimately, this research contributes to growing efforts to harmonize fairness and personalization in LLM-based recommender systems~\cite{lubos2024llm,yang2024behavior,wang2024counterfactual,alves2024user}, contributing to the broader discourse on ethical AI in multimodal and continual learning environments LLM Rec, Behavior Modeling, Counterfactual Fairness, User-Centric Design.


\section{Acknowledgments}
This work was supported by the the Software Engineering Institute of Beihang University and Chinese Government Scholarship. Heartfelt gratitude to Prof. Dr. Xiaoli Lian for invaluable support, supervision, guidance.


{
    \small
    \bibliographystyle{ieeenat_fullname}
    \bibliography{main}
}

\end{document}